\documentclass[conference]{IEEEtran}
\IEEEoverridecommandlockouts
\usepackage{cite}
\usepackage{amsmath,amssymb,amsfonts}
\usepackage{algorithmic}
\usepackage{graphicx}
\usepackage{textcomp}
\usepackage{xcolor}
\usepackage{multirow}
\def\BibTeX{{\rm B\kern-.05em{\sc i\kern-.025em b}\kern-.08em
    T\kern-.1667em\lower.7ex\hbox{E}\kern-.125emX}}
\begin{document}

\title{Sensor-Adaptive Infrared Spectral Reconstruction with Plug-and-Play Diffusion Priors
}
\author{
\IEEEauthorblockN{1\textsuperscript{st} Alireza Siyavashi}
\IEEEauthorblockA{\textit{Chair of Computer Engineering} \\
\textit{Brandenburgische Technische Universit\"at (BTU)} \\
Cottbus, Germany \\
siyavash@b-tu.de}
\and
\IEEEauthorblockN{2\textsuperscript{nd} Jon Schlipf}
\IEEEauthorblockA{\textit{IHP--Leibniz--Institut f\"ur} \\
\textit{Innovative Mikroelektronik} \\
Frankfurt (Oder), Germany \\
schlipf@ihp-microelectronics.com}
\and
\IEEEauthorblockN{3\textsuperscript{rd} Sebastian Reiter}
\IEEEauthorblockA{\textit{Chair of Experimental Physics} \\
\textit{and Functional Materials, BTU} \\
Cottbus, Germany\\
sebastian.reiter@b-tu.de}
\and
\IEEEauthorblockN{4\textsuperscript{th} Inga Fischer}
\IEEEauthorblockA{\textit{Chair of Experimental Physics} \\
\textit{and Functional Materials, BTU} \\
Cottbus, Germany\\
inga.fischer@b-tu.de}
\and
\IEEEauthorblockN{5\textsuperscript{th} Christian Wenger}
\IEEEauthorblockA{\textit{IHP--Leibniz--Institut f\"ur} \\
\textit{Innovative Mikroelektronik} \\
Frankfurt (Oder), Germany \\
wenger@ihp-microelectronics.com}
\and
\IEEEauthorblockN{6\textsuperscript{th} Christian Herglotz}
\IEEEauthorblockA{\textit{Chair of Computer Engineering} \\
\textit{Brandenburgische Technische Universit\"at (BTU)} \\
Cottbus, Germany \\
christian.herglotz@b-tu.de}
}
\IEEEaftertitletext{\vspace{-25pt}}
\maketitle
\begin{abstract}
Hyperspectral sensing enables material identification; however, state-of-the-art spectrometers are costly and bulky, which limits their use in mobile applications.
We address this by proposing sparse spectrum reconstruction from narrow-band photocurrents using a pseudoinverse-guided diffusion model ($\Pi$GDM).
With $\Pi$GDM we use a denoising diffusion probabilistic model (DDPM) to reconstruct the spectrum, which is trained on a large public spectral dataset to learn realistic spectral priors, eliminating the need for paired sensor measurements.
At inference, $\Pi$GDM alternates reverse-diffusion denoising steps with pseudoinverse projection to enforce consistency with measured photocurrents via the calibrated responsivity matrices of sensors.
Consequently, our method is sensor-adaptive: when detector arrays change, we simply substitute the responsivity matrix in the pseudoinverse projection without retraining of the diffusion model.
The resulting computational spectrometer achieves 1.502\% average estimation error, outperforming Tikhonov, Gaussian, compressive-sensing, and multilayer perceptron (MLP) baselines, while providing calibrated uncertainty estimates via Monte Carlo sampling from different random initializations of $\Pi$GDM.
Summarizing, our approach offers an accurate, compact alternative for spectral recovery on resource-constrained platforms.
\end{abstract}
\vspace{-2pt}
\begin{figure*}[t]
  \centering
  \includegraphics[width=\linewidth]{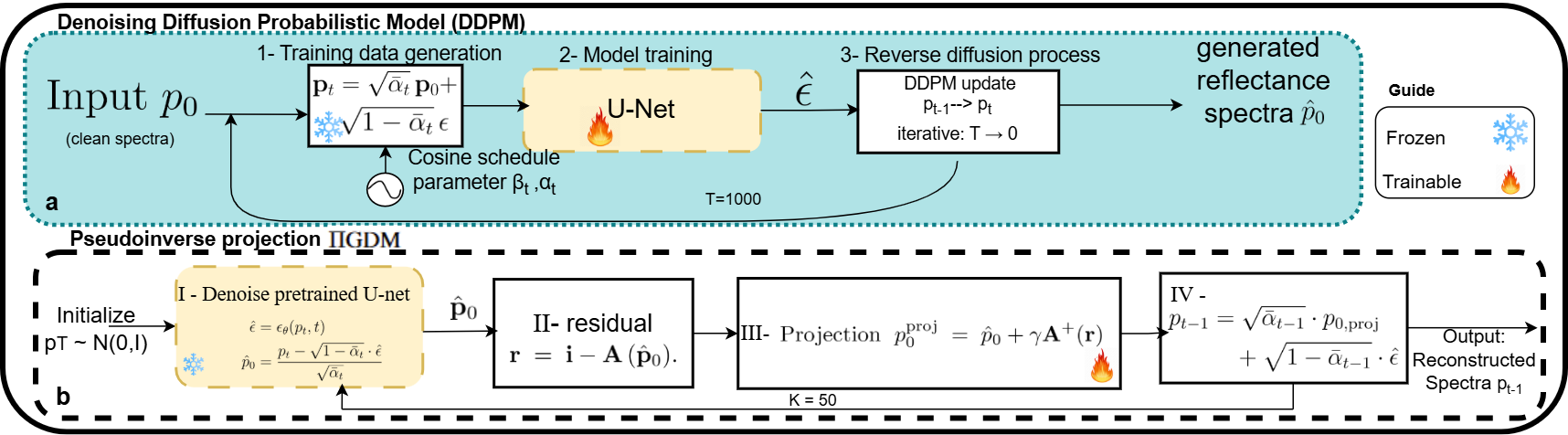}
  \vspace{-0.7cm}
  \caption{Overview of the proposed method. \textbf{a:} Expanded view of the diffusion module separated in three main phases for reflectance spectra generation through an iterative denoising process. \textbf{b:} Full pipeline steps for $\Pi$GDM to reconstruct spectra from trained DDPM.}
  \vspace{-0.7cm}
  \label{fig:method-overview}
\end{figure*}
\vspace{-0.25cm}
\section{Introduction}
\label{sec:intro}
Optical spectroscopy, which analyzes the spectral composition of light entering a spectrometer, can be a valuable tool in a wide variety of applications, including astronomy, agriculture, material analysis, and biomedical detection\cite{taha2025optical,marini2025detecting}. 
Typical spectrometers use diffraction gratings or mechanically modifiable cavities, leading to large, costly, and heavy devices~\cite{gao_computational_2022}.
This limits their applicability, especially in low-cost and mobile applications. 
A way to overcome such limitations is the use of computational spectrometers~\cite{gao_computational_2022}. 
In these methods, instead of directly measuring the spectrum with the spectrometer, an array of sensors~\cite{schlipf_robust_2023} is employed to measure the photocurrent at a few bandwidth-limited wavelengths. 
Then, by exploiting the spectral response of sensors, reconstruction of the entire spectrum becomes possible. 
These spectrum reconstruction methods are classified into traditional reconstruction algorithms and deep learning-based models~\cite {xue_advances_2024}.

Classical methods, such as the Tikhonov regularization and Compressive Sensing (CS)~\cite{yuan_wavelength-scale_2021,kurokawa_filter-based_2011,ma_robust_2024}, are computationally lightweight.
However, their hyperparameters need to be tuned when the process of sensor manufacturing changes, which leads to different sensor responsivity.
Schlipf et al. \cite{schlipf_robust_2023} reconstructed spectra using the photocurrent of metasurface photodetectors by solving a linear inverse problem using Gaussian basis functions and matrix inversion. 
They assumed that the input optical spectrum can be represented as a superposition of Gaussian basis functions.
However, this method suffers from limited reconstruction accuracy due to the limited number of detectors and the simplified assumption of Gaussian spectral bases.
As their coefficients are recovered by direct matrix inversion, the reconstruction can become sensitive to measurement noise and overlapping detector responses.

On the other hand, end-to-end deep learning methods (e.g., MLP) rely on large datasets, which can be a major practical bottleneck when collecting paired sensor readouts and ground-truth spectra.
Additionally, these methods may suffer from limited generalizability and biases in the case of uneven distribution of samples~\cite{zhang_survey_2022}.
A practical way to alleviate this bottleneck is to decouple learning from data collection.
Rather than learning an end-to-end mapping from photocurrents to spectra using paired measurements, we train a DDPM on a large, publicly available spectral library dataset without requiring any sensor data.
This decoupling offers two key advantages: (1) we can leverage abundant existing spectral databases comprising thousands of diverse material reflectances, which is the target application of our spectral reconstruction algorithm, and (2) the same trained model generalizes across different sensor configurations without retraining.
At inference, measurement consistency is enforced by projecting diffusion-generated spectra onto the subspace that matches observed photocurrents via the calibrated forward model~\cite{schlipf_robust_2023}, which calculates from sensors' responsivity matrices and the generated spectra.

Existing Plug-and-Play methods~\cite{zheng_deep_2021}, which separate the learned spectral prior from the sensor-specific measurement model, often rely on MLP-generated priors for spectral reconstruction~\cite{zhang_snapshot_2024}. 
Zheng et al. note that such generator prior PnP approaches can generalize poorly to new sensors due to dataset bias, be sensitive to initialization, and mismatch real, model-induced artifacts when the denoiser is trained only for additive Gaussian noise~\cite{zheng_deep_2021}. 
Moreover, MLP generator priors are typically \emph{deterministic}: for a given photocurrent measurement, they return a single spectrum and do not naturally provide uncertainty intervals. 
This is a practical limitation for sparse narrow-band sensing, where multiple spectra may explain the same measurement; uncertainty bands help reveal ambiguity and prevent overconfident material identification.

In this paper, we trained the DDPM, which is used for generating synthetic data that captures the distribution of realistic spectra from a large public dataset, intending to learn a sensor-adaptive spectral prior that can be applied to any calibrated detector array without requiring retraining.
In the spectral reconstruction step, each run of $\Pi$GDM~\cite{song2023pseudoinverse} starts from random noise (with the same dimensionality as the target spectrum) and iteratively refines it into a reconstructed spectrum that matches the measured photocurrents. 
The pre-trained DDPM serves as a learned \emph{spectral prior}, steering the reconstruction toward spectra that are realistic with respect to the public spectral dataset used for training. 
After each denoising step, $\Pi$GDM applies a physics-based correction using the pseudoinverse of the sensors\ responsivity matrix rule to control whether the measured photocurrents agree with the predicted photocurrents. 
Repeating the reconstruction with different noise initializations yields multiple \emph{measurement-conditioned reconstructions}, i.e., different spectra that all explain the same photocurrent measurement.
 
To the best of our knowledge, this is the first use of $\Pi$GDM for infrared spectrum reconstruction in the $nm$ range and silicon–germanium (SiGe) photocurrent sensors with the modified DDPM.
The method combines the generality of pretrained DDPM~\cite{ho_denoising_2020} with explicit physical constraints and avoids retraining for new set of sensors. 
Applied to metasurface photocurrent measurements, it yields high-quality spectra with a sensor-adaptive prior.

The contribution of our paper is summarized to:
\begin{enumerate}
\setlength{\itemsep}{0pt}
\setlength{\topsep}{0pt}
\setlength{\parsep}{0pt}
\setlength{\partopsep}{0pt}
    \item \textbf{Generalized DDPM for spectra generation} We introduce a light\-weight neural diffusion model that learns the structure of real spectra from public datasets.
    \item \textbf{PnP with pseudoinverse guidance:} integrating a pretrained DDPM into the spectral reconstruction process with physics-based measurement corrections for sensor-adaptive reconstruction.
    \item \textbf{Calibrated posterior uncertainty:}
     Provides confidence bands per wavelength with empirically validated coverage (90\% intervals contain ground truth 90\% of the time), showing reliability beyond Root Mean Square Error (RMSE) and Mean Absolute Percentage Error (MAPE).
     \item \textbf{Benchmarking:}
     Comparison and evaluation with multiple benchmarking methods.
\end{enumerate}
The remainder of the paper is organized as follows.
In Section~\ref{sec:premilinaries} we introduce the physical law and equation of the photodetector.
We also explain two approaches that we utilize for spectra reconstruction from the literature.
Section~\ref{sec:method} delineates our proposed model architecture.
Section~\ref{sec:results&evaluation} describes the evaluation protocol and metrics, and reports quantitative and qualitative comparisons against the baselines.
Section~\ref{sec:conclusion} is an overview of our results and future work.
\section{preliminaries}
\label{sec:premilinaries}
\subsection{Physical Model of Photocurrent Sensors}
\label{sec:forward model}
We model the $m$-channel photodetector where $m$ is the number of sensors as a linear spectral mixer sampled on a uniform spectrum $\{\lambda_j\}_{j=1}^n$ with $m$ = 9 and $n$ = 301 wavelength, $\lambda_1=1300$\,nm, $\lambda_n=1600$\,nm, and a wavelength step of $\Delta\lambda=1$\,nm. 
We adopt a physical model (called forward model) to calculate the photocurrents from the photodetector $\boldsymbol{i}$ as
\vspace{-0.25cm}
\begin{equation}\label{eq:forward}
\boldsymbol{i}=\boldsymbol{R}\mathbf{p}+\boldsymbol{\eta},
\end{equation}
where $\boldsymbol{i}\in\mathbb{R}^m$ are photocurrents [A], $\mathbf{p}\in\mathbb{R}^n$ is the discretized power spectrum [W/nm] spectra (which we target to reconstruct), $\boldsymbol{\eta}$ accounts for sensor noise and model mismatch, which is additive Gaussian noise, and $\boldsymbol{R}\in\mathbb{R}^{m\times n}$ stacks the calibrated responsivities of the sensors on the photodetector [A/W]~\cite{schlipf_robust_2023}.
\subsection{Nonnegative Linear Regularization}
\label{sec:first method}
Given the forward model in Equation \eqref{eq:forward}, reconstructing the reflectance spectrum
$\mathbf{p}\!\in\!\mathbb{R}^{n}$ from $m$ photocurrents is underdetermined when $m\!<\! n$
(e.g., few detectors, many wavelengths). A common solution to calculate the estimated $\mathbf{\hat{p}}$ is the Tikhonov-regularized nonnegative least squares (NNLS) \cite{xue_advances_2024}:
\vspace{-0.25cm}
\begin{equation}
\min_{\mathbf{\hat{p}}\ge 0}\; \tfrac12\lVert \boldsymbol{R}\mathbf{\hat{p}}-\boldsymbol{i}\rVert_2^2
\;+\; \tfrac{\Lambda}{2}\,\lVert \mathbf{D}\mathbf{\hat{p}}\rVert_2^2,
\label{eq:tikhonov}
\vspace{-0.15cm}
\end{equation}
where $\mathbf{D}$ is the second-derivative operator (penalizing the spectral curvature) to encourage smooth
spectra.
This model has low computational complexity and is stable, but quadratic smoothing can
oversuppress narrow spectral features and requires tuning the hyperparameter $\Lambda$, which sets the importance of smoothness.

\subsection{Sparse Compressive-sensing Priors}
\label{sec:second method}
As a second benchmark model, we modeled the spectrum $\mathbf{p}\in\mathbb{R}^n$ as sparse in a discrete cosine transform (DCT-II)~\cite{candes2006robust} dictionary
$\boldsymbol{\Phi}\in\mathbb{R}^{n\times k}$ (first $k$ atoms), so that $\mathbf{p}\approx \boldsymbol{\Phi}\boldsymbol{x}$
with coefficients $\boldsymbol{x}\in\mathbb{R}^k$. Let $\boldsymbol{B}:=\boldsymbol{R}\boldsymbol{\Phi}$ denote the forward
operator from coefficients to photocurrents $\boldsymbol{i}$. We estimate $\boldsymbol{x}$ by Elastic-Net–regularized
least squares~\cite{candes2006robust,xue_advances_2024}:
\begin{equation}
\min_{\boldsymbol{x}}\; \tfrac12\|\boldsymbol{B}\boldsymbol{x}-\boldsymbol{i}\|_2^2
\;+\;\Lambda\!\left(\rho\|\boldsymbol{x}\|_1+(1-\rho)\|\boldsymbol{x}\|_2^2\right),
\label{eq:enet}
\end{equation}
where $\Lambda =1$ controls the regularization strength and $\rho\in[0,1]$ mixes the $\ell_1/\ell_2$ penalties that balance sparsity. 
We select $\Lambda$ by grid search over $[10^{-3},10^{1}]$ (25 log-spaced values) on the training set, and fix $\rho=0.95$ as the value minimizing validation RMSE. 
The reconstruction is $\hat{\mathbf{p}}=\max(\boldsymbol{\Phi}\hat{\boldsymbol{x}},0)$.
\section{Proposed reconstruction method}
\label{sec:method}
In this section, we explain how we reconstruct spectra \(\mathbf{p} \in [0,1]^n\) from measured photocurrents $\boldsymbol{i}$ using $\Pi$GDM, which couples a learned spectral prior (modeled by DDPM that learns the distribution of clean spectra through noise prediction) with responsivity-based projection via the forward model (Eq.~\ref{eq:forward}), as illustrated in Fig.~\ref{fig:method-overview}. 

The DDPM pipeline, which denoises noisy input spectra (Fig.~\ref{fig:method-overview}a) consists of three phases: 
First, we define the forward diffusion process (Equation~\eqref{eq:forward_diffusion}) that progressively adds Gaussian noise to clean spectra from public datasets, creating training pairs $(\boldsymbol{p}_t, \boldsymbol{\epsilon})$ at various timesteps $t \in \{1, \ldots, T\}$. 
Second, we train a U-Net to predict the added noise $\boldsymbol{\epsilon}$ from any noisy spectrum $\boldsymbol{p}_t$ and the timestep $t$, enabling the model to separate noise from spectra. 
Third, at inference, starting from pure noise $\boldsymbol{p}_T\sim\mathcal{N}(\mathbf{0},\mathbf{I})$, the reverse diffusion process iteratively applies the trained U-Net to predict and remove noise at each timestep, progressively recovering a realistic spectrum.
Once trained on public spectral datasets, the U-Net model is frozen and reused for reconstruction within the DDPM framework.

In inference (Fig.~\ref{fig:method-overview}b), $\Pi$GDM iteratively refines random noise into an estimated spectrum by alternating DDPM denoising with pseudoinverse projection to enforce measurement consistency (Section~\ref{sub:pigdm}). 
This yields the final reconstructed spectrum $\hat{\mathbf{p}} \in [0,1]^n$.
Crucially, our framework supports \emph{zero-shot} reconstruction: the same pretrained DDPM applies to any calibrated sensor by simply substituting the responsivity matrix $\boldsymbol{R}$, without retraining.
\subsection{Noisy Data Generation for Training}
We define the forward diffusion process with \(T = 1000\) timesteps, as illustrated in phase 1 of Fig.~\ref{fig:method-overview}a. Let \(\boldsymbol{p}_0\in\mathbb{R}^n\) denote a clean (normalized) reflectance spectrum from the public training datasets, and \(\boldsymbol{p}_t\) its noisy version at timestep \(t\in\{1,\dots,T\}\). Given i.i.d. Gaussian noise \(\boldsymbol{\epsilon}\sim\mathcal{N}(\mathbf{0},\mathbf{I}_n)\), the noisy spectrum at any timestep \(t\) can be computed directly as:
\vspace{-0.25cm}
\begin{equation}
\boldsymbol{p}_t=\sqrt{\bar{\alpha}_t}\,\boldsymbol{p}_0+\sqrt{1-\bar{\alpha}_t}\,\boldsymbol{\epsilon}
\label{eq:forward_diffusion}
\end{equation}
where \(\bar{\alpha}_t\in(0,1]\) controls the signal-to-noise ratio at timestep \(t\). We adopt the cosine noise schedule~\cite{dhariwal_diffusion_2021} with offset \(s=0.008\):
\begin{equation}
\begin{aligned}
\bar{\alpha}_t &\propto 
\cos^2\!\left(\frac{t/T + s}{1+s}\cdot\frac{\pi}{2}\right),\quad
\beta_t = \min\!\left(0.999,\; 1 - \frac{\bar{\alpha}_t}{\bar{\alpha}_{t-1}}\right),\\
\alpha_t &= 1-\beta_t,\quad
\bar{\alpha}_t = \prod_{j=1}^{t}\alpha_j.
\end{aligned}
\vspace{-0.15cm}
\label{eq:cosine_schedule}
\end{equation}
This schedule gradually increases noise while preserving spectral structure longer than a linear schedule~\cite{ho_denoising_2020}, leading to more stable training and higher-quality reconstructions. During training, we randomly sample a timestep \(t\) for each clean spectrum and generate the corresponding noisy version \(p_t\) on-the-fly using Eq.~\eqref{eq:forward_diffusion}.
\subsection{U-Net Architecture with Time Conditioning}
We employ a 1D U-Net denoising network $\boldsymbol{\epsilon}_{\theta}(\boldsymbol{p}_t, t)$ as the backbone~\cite{ho_denoising_2020}. 
The network operates on spectral sequences $\boldsymbol{p}_t \in \mathbb{R}^n$ ($n=301$ wavelengths), outputting the predicted noise $\hat{\boldsymbol{\epsilon}} \in \mathbb{R}^n$.
Temporal conditioning is implemented via sinusoidal embeddings~\cite{vaswani2017attention} processed through a small MLP and injected into each residual block of U-net, enabling timestep-adaptive denoising across diffusion steps.
The architecture follows with four resolution levels (channel multipliers $1\times, 2\times, 4\times, 8\times$ over a base of 96 channels), two residual blocks per level, and a bottleneck of two time-conditioned residual blocks~\cite{he2016deep}. Each residual block uses $3\times1$ convolutions with padding, group normalization, SiLU~\cite{ho_denoising_2020} activations, and 5\% dropout. 
Downsampling uses strided convolutions; upsampling uses nearest-neighbor interpolation followed by convolution. 
Skip connections concatenate encoder and decoder features at matching resolutions. 
The model totals 4.8M parameters.

The training objective minimizes the mean squared error between predicted and true noise:
\begin{equation}
\mathcal{L}(\theta) = \mathbb{E}_{\boldsymbol{p}_0, \boldsymbol{\epsilon}, t} \left[ \|\boldsymbol{\epsilon} - \boldsymbol{\epsilon}_{\theta}(\boldsymbol{p}_t, t)\|^2_2 \right].
\label{eq:ddpm_loss}
\end{equation}
This trains the model to denoise from any corruption level $t$, providing a flexible spectral prior for reconstruction. From the predicted noise $\hat{\boldsymbol{\epsilon}}=\boldsymbol{\epsilon}_{\theta}(\boldsymbol{p}_t, t)$, the clean spectrum estimate is recovered as $\hat{\boldsymbol{p}}_0 = (p_t - \sqrt{1-\bar{\alpha}_t}\,\hat{\boldsymbol{\epsilon}}) \,/\, \sqrt{\bar{\alpha}_t}$.
\subsection{Pseudoinverse-Guided Diffusion Sampling}
\label{sub:pigdm}
At inference (Fig.~\ref{fig:method-overview}b), the trained U-Net is coupled with the responsivity matrix of sensors projection to enforce measurement consistency. 
We reformulate the forward model as $A(\boldsymbol{p}) = \tfrac{1}{2}\boldsymbol{R}\boldsymbol{p} + \tfrac{1}{2}\boldsymbol{R}\mathbf{1}_n$, which maps DDPM-normalized reflectance $\boldsymbol{p}\in[-1,1]^n$ to photocurrents ($\mathbf{p}$).
Starting from pure noise $\boldsymbol{p}_T \sim \mathcal{N}(0, I)$, we iteratively refine the spectrum through Steps I--IV:

\textbf{Step I (Denoise):} The pretrained U-Net predicts noise $\hat{\boldsymbol{\epsilon}} = \boldsymbol{\epsilon}_{\theta}(\boldsymbol{p}_t, t)$ and proposes an initial clean spectrum estimate $\hat{\boldsymbol{p}}_0$.
\textbf{Step II (Residual):} Compute the measurement residual $\boldsymbol{r} = \boldsymbol{i} - A(\hat{\boldsymbol{p}}_0)$.

\textbf{Step III (Projection):} Apply pseudoinverse-based correction to ensure consistency with measured photocurrents:
\vspace{-0.25cm}
\begin{equation}
\mathbf{p}_{0,\mathrm{proj}} = \hat{\boldsymbol{p}}_0 + \gamma A^{+}(\boldsymbol{r}),
\label{eq:projection}
\vspace{-0.25cm}
\end{equation}
where $A^{+}$ is the Moore--Penrose pseudoinverse~\cite{barata2012moore} of $A$ and $\gamma \in [0,1]$ controls projection strength. We fix $\gamma=1.0$, the value minimizing validation RMSE.

\textbf{Step IV (DDPM Step):} Advance to the next timestep by computing $\boldsymbol{p}_{t-1}$ from $\mathbf{p}_{0,\mathrm{proj}}$ using the standard DDPM update rule utilizing Equation~\eqref{eq:forward_diffusion}~\cite{ho_denoising_2020}.
$\boldsymbol{p}_{t-1}$ is then used as the input to step I in the next iteration.
This loop repeats for $K=50$ uniformly spaced steps from $t=T$ to $t=0$ (a standard DDIM-style acceleration~\cite{ho_denoising_2020} that sub-samples the full $T=1000$ trajectory), yielding the final reconstructed spectrum $\hat{\mathbf{p}}\in[0,1]^n$.
\begin{table}[t]
\centering
\caption{Reconstruction errors across six reflectance spectra samples.}
\label{tab:mape_comparison}
\normalsize
\setlength{\tabcolsep}{2.6pt}      
\renewcommand{\arraystretch}{1.12} 
\resizebox{\columnwidth}{!}{%
\begin{tabular}{l cc cc cc cc cc}
\hline\hline
\multirow{2}{*}{Samples}
  & \multicolumn{2}{c}{Tikhonov}
  & \multicolumn{2}{c}{Gaussian}
  & \multicolumn{2}{c}{Compressive Sensing}
  & \multicolumn{2}{c}{MLP Priors}
  & \multicolumn{2}{c}{Diffusion ($\Pi$GDM)} \\
& RMSE & MAPE & RMSE & MAPE & RMSE & MAPE & RMSE & MAPE & RMSE & MAPE \\
\hline
p1   & 17.656 & 18.023 & 13.856 & 11.492 & 3.348 & 2.903 & 10.720 & 10.231 & 1.854 & 1.454 \\
p2   & 5.968  & 7.047  & 15.259 & 11.506 & 3.012 & 2.363 & 9.302  & 9.122  & 2.708 & 1.972 \\
p3   & 12.400 & 13.731 & 13.819 & 11.648 & 3.245 & 2.628 & 10.145 & 10.002 & 1.700 & 1.391 \\
p4   & 10.285 & 9.842  & 16.534 & 11.114 & 1.236 & 0.848 & 11.000 & 10.860 & 2.625 & 1.431 \\
p5   & 12.617 & 14.385 & 12.188 & 11.748 & 1.530 & 1.672 & 9.335  & 9.230  & 1.298 & 1.515 \\
p6   & 19.382 & 12.190 & 17.805 & 10.942 & \textbf{0.250} & \textbf{0.249} & 9.650 & 9.357 & 2.483 & 1.254 \\
\hline
Mean & 13.052 & 12.536 & 14.910 & 11.408 & 2.115 & 1.776 & 9.960 & 9.800 & \textbf{2.111} & \textbf{1.502} \\
\hline\hline
\end{tabular}
}
\end{table}
\vspace{-0.2cm}
\section{evaluation}
\label{sec:results&evaluation} 
For the generative prior models (DDPM and MLP), we use the datasets USGS v7~\cite{raymond_kokaly_usgs_2017}, SPECCHIO~\cite{hueni_spectral_2009} (1429--2500~nm), and ECOSTRESS~\cite{meerdink_ecostress_2019}. 
We evaluated our model using public and private spectra.
The missing wavelengths in SPECCHIO are filled by the means of USGS + EcoSTRESS.
Baseline hyperparameters (e.g., the regularization weight $\lambda$ for Tikhonov and compressive sensing) are tuned on the same validation split used to select $\Pi$GDM's projection weight $\gamma$.
For the MLP prior, we use the architecture and training method of \cite{zhang_snapshot_2024}, reporting the best validation checkpoint from all epochs.

Our target test set includes 2 SiGe arrays of sensor channels~\cite{schlipf_robust_2023}. 
Sensor data comprises nine photodetector-calibrated responsivities within 1300--1600~nm and measured reflectance spectra (P1--P6) plus soil, PVC and PET. 
Ground-truth spectra and photocurrents are used to assess the reconstruction accuracy.
Accuracy is measured against the ground-truth $\mathbf{p}$ using RMSE and MAPE:
\begin{equation}
\label{eq:rmse_mape_inline}
\mathrm{RMSE} = \sqrt{\frac{1}{N}\sum_{j=1}^{n}(\hat{p}_j - p_j)^2},
\mathrm{MAPE} = \frac{100}{N}\sum_{j=1}^{n}\left|\frac{\hat{p}_j - p_j}{p_j}\right|.
\end{equation}
where $j$ is wavelength index and $N$ is the number of reflectance spectra samples.
We train a 1D DDPM prior on publicly available reflectance spectra (1300–1600 nm) using a standard training and validation split; no sensor measurements are used during the DDPM training.
At test time, we performed a zero-shot reconstruction by plugging another measured independent calibrated responsivity $\boldsymbol{R}_2$ and photocurrents $\boldsymbol{i}$ into the $\Pi$GDM model. 
\begin{figure}[t]
    \centering
    \includegraphics[width=0.48\textwidth]{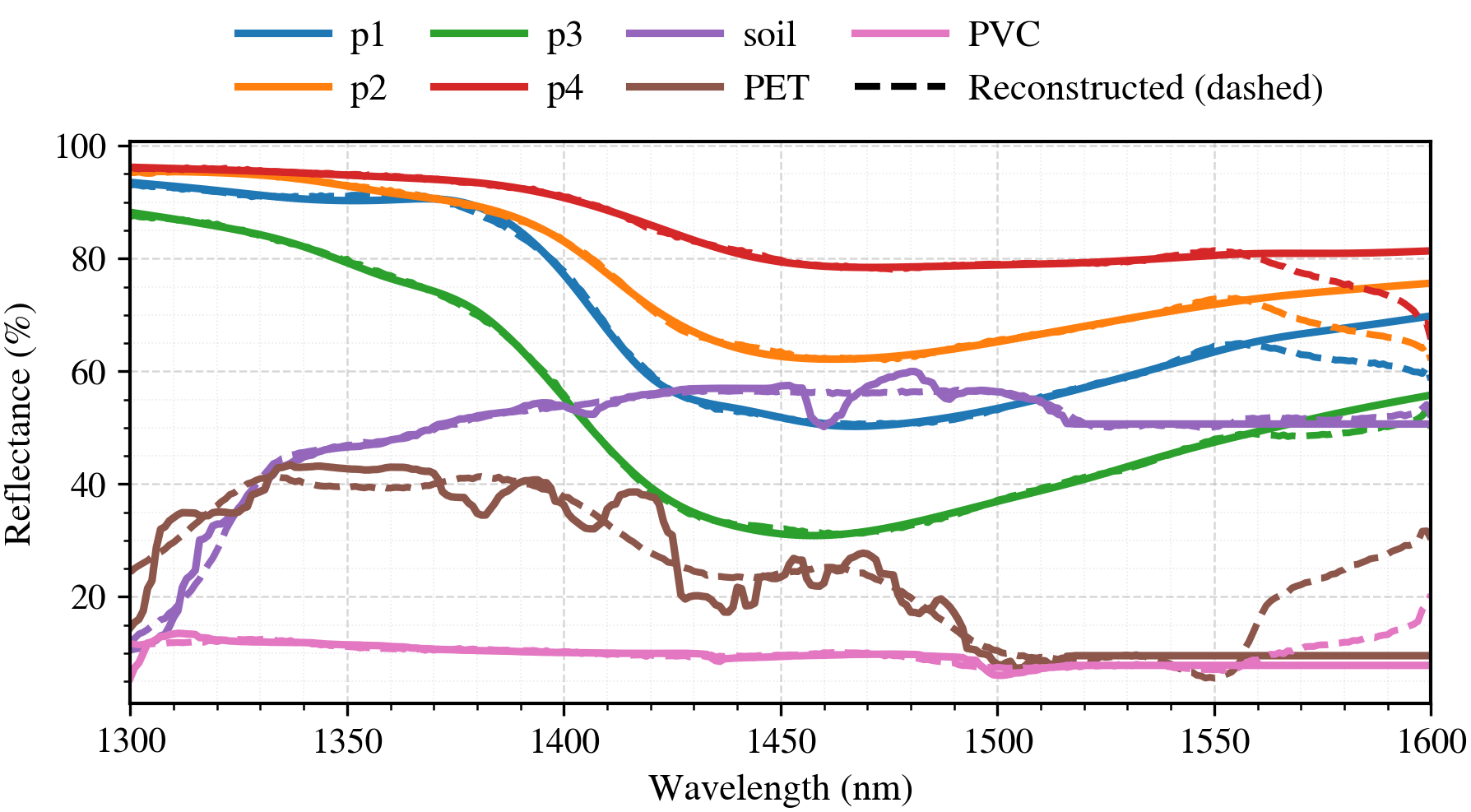}
    \vspace{-0.3cm}
    \caption{Reflectance spectra reconstruction with $\Pi$GDM To test sensor adaptivity, we replace the second independently calibrated responsivity matrix $\boldsymbol{R}_{2}$ (nine channels) corresponding to a different sensor array, while keeping the pretrained diffusion prior fixed.
}
    \vspace{-0.5cm}
    \label{fig:reflectance_reconstruction}
\end{figure}
Next to the accuracy evaluation, we calculate Uncertainty by running the projection of $\Pi$GDM $S=128$ times with independent Gaussian initializations, i.e., $\boldsymbol{p}_{t_1}^{(s)} \sim \mathcal{N}(0,I)$ for $s=1,\dots,S$ yielding spectra $\boldsymbol{p}^{(s)}\!\in\!\mathbb{R}^n$ consistent with the measurements and learned prior. 
The set $\{\mathbf{P}^{(s)}\}_{s=1}^S$ approximates $\mathbf{P}(\hat{\boldsymbol{p}}\mid \boldsymbol{i},\boldsymbol{R})$, the distribution of unknown spectrum $\hat{\boldsymbol{p}}$ given measured photocurrent $\boldsymbol{i}$ and responsivity $\boldsymbol{R}$. 
We report the credible intervals of the posterior mean $\bar{\mathbf{P}}(\lambda)=\frac{1}{S}\sum_{s=1}^S \mathbf{P}^{(s)}(\lambda)$ and pointwise $90\%$ credible intervals (5\textsuperscript{th}–95\textsuperscript{th} percentiles over $s$). 
\begin{figure}[t]
  \centering
  \includegraphics[width=\linewidth]{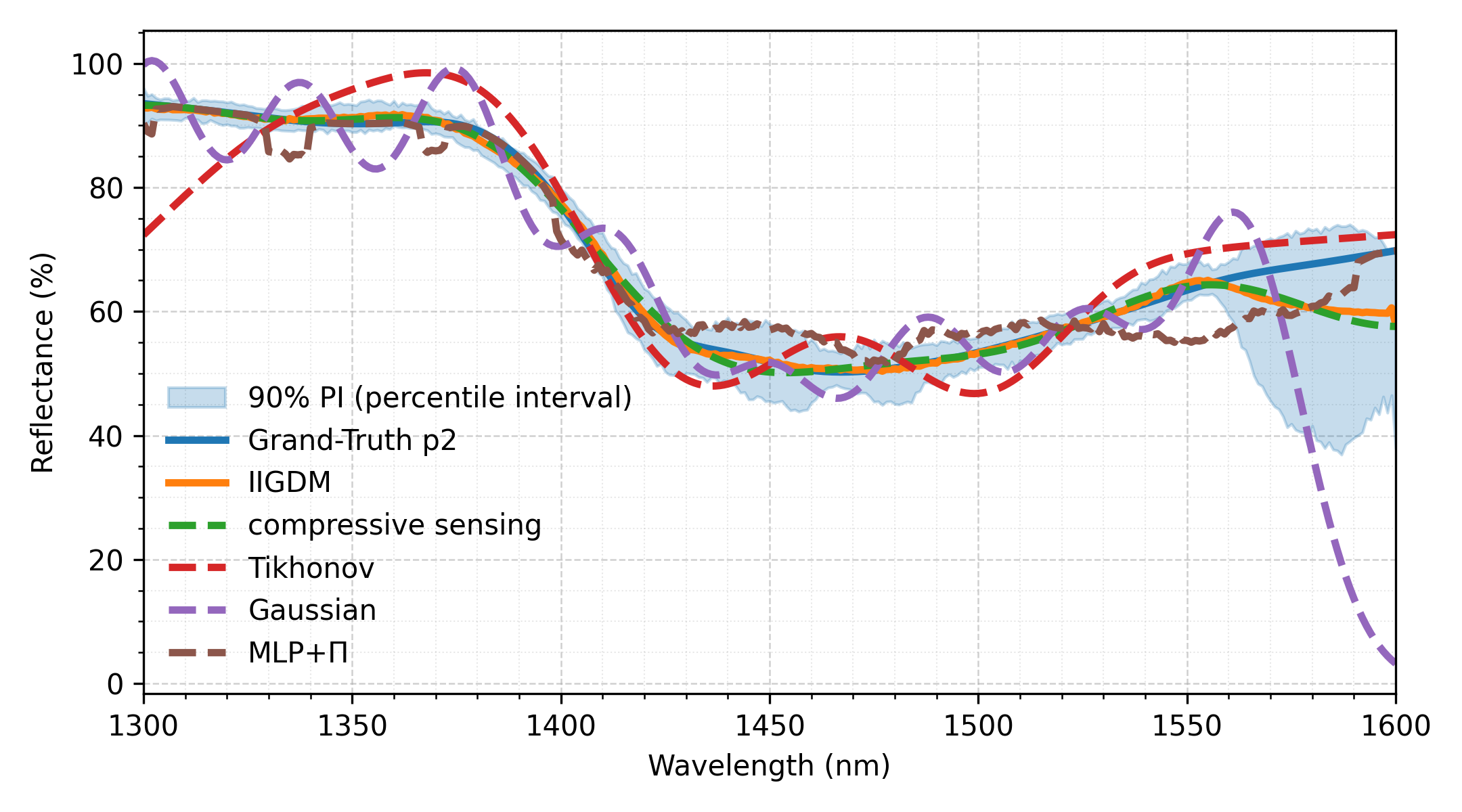}
  \vspace{-0.5cm}
  \caption{Comparison of NIR reflectance reconstructions: $\Pi$GDM (orange; shaded 90\% percentile interval) versus ground truth (blue), along with traditional baselines (compressive sensing, Tikhonov, Gaussian) and a learning-based prior (MLP+$\Pi$).
}
\vspace{-0.7cm}
  \label{fig:recon_overlay}
\end{figure}

Figure~\ref{fig:reflectance_reconstruction} compares the reconstructed reflectance spectra with the ground truth when using a responsivity matrix different from the one used during training.
We can see that the diffusion-based model (DDPM prior with $\Pi$GDM) recovers the global shape and characteristic across the 1300–1600\,nm band.
Table~\ref{tab:mape_comparison} reports RMSE and MAPE for six reflectance spectra (p1--p6) against baselines and MLP prior. 
Overall, our diffusion-based $\Pi$GDM achieves the lowest average error (RMSE $2.111$, MAPE $1.502\%$). 
Compressive sensing achieves competitive performance and can even outperform $\Pi$GDM on some samples; however, its reconstruction quality strongly depends on the regularization parameter $\alpha$, which typically needs to be retuned for each new calibrated sensor set. 
In contrast, $\Pi$GDM is trained once on reflectance spectra and incorporates the sensor responsivity only at inference via the projection step, enabling plug-and-play deployment across sensors without retraining or retuning.
We demonstrate this by swapping calibrated responsivity matrices and comparing both the ground truth and reconstruction results in Figs.~\ref{fig:reflectance_reconstruction} and~\ref{fig:recon_overlay}.

The MLP prior also avoids per-sensor tuning, but it yields substantially higher errors than $\Pi$GDM, indicating that the diffusion prior captures the spectral manifold more effectively.

Figure~\ref{fig:recon_overlay} compares the proposed method with classical and learning-based baselines and visualizes the associated \emph{uncertainty region}, shown as the shaded area between the 5th and 95th percentiles of the posterior samples $\{\mathbf{P}^{(s)}(\lambda)\}_{s=1}^S$ at each wavelength. 
The uncertainty bands are narrow in wavelength regions where the sensor provides strong information, i.e., where the responsivity matrix $\boldsymbol{R}$ is well conditioned with high sensitivity and low inter-channel collinearity. 
Conversely, the bands widen in \emph{poorly constrained} regions, particularly in the long-wavelength tail (1550--1600\,nm), where the sensor sensitivity diminishes. 
Across all scenes, the ground-truth spectra remain within the 90\% uncertainty intervals, indicating well-calibrated posterior uncertainty.
\section{Conclusion}
\label{sec:conclusion}
We propose a model for spectral reconstruction using sparse photocurrent sensors, yielding average MAPE = $1.502\%$, trained on a broad dataset and usable with calibrated photocurrent sensors by simply plugging in the sensor’s responsivity without retraining.
The proposed diffusion model is sensor-agnostic (independent in terms of manufacturing process and characteristic), and can reconstruct the general shape of spectra.
The model returns calibrated probabilities and per-wavelength uncertainty bands, enabling decision-making under ambiguity. 
When confidence is low, these uncertainty estimates indicate spectra ranges in which additional measurements will be most informative.
In future work, we are going to deploy this method for sensor arrays for material classification by drones. 
\vspace{-0.45cm}
\section{acknowledgement}
The authors would like to thank the Federal Ministry of Research, Technology and Space of Germany (BMFTR) for financial support under the Innovations campus Elektronik und Mikrosensorik – iCampus Cottbus“ project, grant numbers 16ME0424 and 16ES1131.

\bibliographystyle{IEEEtran}
\bibliography{refs}

@article{schlipf_robust_2023,
  title   = {Robust Si/Ge heterostructure metasurfaces as building blocks for wavelength-selective photodetectors},
  author  = {Schlipf, J. and Berkmann, F. and Yamamoto, Y. and Reichenbach, M. and Veleski, M. and Kawaguchi, Y. and Mörz, F. and Tomm, J. W. and Weißhaupt, D. and Fischer, I. A.},
  date    = {2023-03-20},
  journal = {Applied Physics Letters},
  volume  = {122},
  number  = {12},
  pages   = {121701},
  doi     = {10.1063/5.0134458},
  issn    = {0003-6951, 1077-3118},
  urldate = {2025-04-25},
  langid  = {english},
    year = 2023
}

@article{zhang_survey_2022,
	title = {A survey on computational spectral reconstruction methods from {RGB} to hyperspectral imaging},
	volume = {12},
	issn = {2045-2322},
	doi = {10.1038/s41598-022-16223-1},
	pages = {11905},
	number = {1},
	journal = {Scientific Reports},
	shortjournal = {Sci Rep},
	author = {Zhang, J. and Su, R. and Fu, Q. and Ren, W. and Heide, F. and Nie, Y.},
	urldate = {2025-08-18},
	date = {2022-07-13},
	langid = {english},
    year = 2022
}

@article{ho_denoising_2020,
  title={Denoising diffusion probabilistic models},
  author={Ho, J. and Jain, A. and Abbeel, P.},
  journal={Advances in neural information processing systems},
  volume={33},
  pages={6840--6851},
  year={2020}
}

@article{dhariwal_diffusion_2021,
  title={Diffusion models beat gans on image synthesis},
  author={Dhariwal, P. and Nichol, A.},
  journal={Advances in neural information processing systems},
  volume={34},
  pages={8780--8794},
  year={2021}
}

@misc{raymond_kokaly_usgs_2017,
	title = {{USGS} Spectral Library Version 7 Data},
	rights = {Creative Commons Zero v1.0 Universal},
	doi = {10.5066/F7RR1WDJ},
	publisher = {U.S. Geological Survey},
	author = {Kokaly, R. and Clark, R. N. and Swayze, G. A. and Livo, K. E. and Hoefen, T. M. and Pearson, N. C. and Wise, R. A. and Benzel, W. M. and Lowers, H. A. and Driscoll, R. L. and Klein, A. J.},
	urldate = {2025-08-19},
	date = {2017},
}

@article{hueni_spectral_2009,
  title        = {The spectral database {SPECCHIO} for improved long-term usability and data sharing},
  volume       = {35},
  rights       = {https://www.elsevier.com/tdm/userlicense/1.0/},
  issn         = {00983004},
  url          = {https://doi.org/10.1016/j.cageo.2008.03.015},
  doi          = {10.1016/j.cageo.2008.03.015},
  pages        = {557--565},
  number       = {3},
  journaltitle = {Computers \& Geosciences},
  shortjournal = {Computers \& Geosciences},
  journal      = {Computers \& Geosciences},
  author       = {Hueni, A. and Nieke, J. and Schopfer, J. and Kneubühler, M. and Itten, K. I.},
  urldate      = {2025-09-14},
  date         = {2009-03},
  month        = mar,
  year         = {2009},
  langid       = {english},
}

@article{vaswani2017attention,
  title={Attention is all you need},
  author={Vaswani, A. and Shazeer, N. and Parmar, N. and Uszkoreit, J. and Jones, L. and Gomez, A. N. and Kaiser, {\L}. and Polosukhin, I.},
  journal={Advances in neural information processing systems},
  volume={30},
  year={2017}
}

@inproceedings{he2016deep,
  author    = {He, K. and Zhang, X. and Ren, S. and Sun, J.},
  title     = {Deep {R}esidual {L}earning for {I}mage {R}ecognition},
  booktitle = {Proc. IEEE Conf. on Computer Vision and Pattern Recognition (CVPR)},
  year      = {2016},
  pages     = {770--778},
  address   = {Las Vegas, NV, USA},
  month     = {June},
  publisher = {IEEE},
  doi       = {10.1109/CVPR.2016.90}
}

@article{taha2025optical,
  title={Optical Spectroscopy of Cerebral Blood Flow for Tissue Interrogation in Ischemic Stroke Diagnosis},
  author={Taha, B. A. and Kadhim, A. C. and Addie, A. J. and Al-Jubouri, Q. and Azzahrani, A. S. and Haider, A. J. and Alkawaz, A. N. and Arsad, N.},
  journal={ACS Chemical Neuroscience},
  volume={16},
  number={5},
  pages={895--907},
  year={2025},
  publisher={ACS Publications}
}

@article{marini2025detecting,
  title={Detecting clusters and groups of galaxies populating the local Universe in large optical spectroscopic surveys},
  author={Marini, I. and Popesso, P. and Dolag, K. and Bravo, M. and Robotham, A. and Tempel, E. and Li, Q. and Yang, X. and Csizi, B. and Behroozi, P. and others},
  journal={Astronomy \& Astrophysics},
  volume={694},
  pages={A207},
  year={2025},
  publisher={EDP Sciences}
}

@article{gao_computational_2022,
  title={Computational spectrometers enabled by nanophotonics and deep learning},
  author={Gao, L. and Qu, Y. and Wang, L. and Yu, Z.},
  journal={Nanophotonics},
  volume={11},
  number={11},
  pages={2507--2529},
  year={2022},
  publisher={De Gruyter}
}

@article{xue_advances_2024,
	title = {Advances in Miniaturized Computational Spectrometers},
	volume = {11},
	rights = {© 2024 The Author(s). Advanced Science published by Wiley-{VCH} {GmbH}},
	issn = {2198-3844},
        year = 2024,
	doi = {10.1002/advs.202404448},
	pages = {2404448},
	number = {47},
	journal = {Advanced Science},
	author = {Xue, Q. and Yang, Y. and Ma, W. and Zhang, H. and Zhang, D. and Lan, X. and Gao, L. and Zhang, J. and Tang, J.},
	urldate = {2025-08-25},
	date = {2024},
	langid = {english},
	}

@article{kurokawa_filter-based_2011,
	title = {Filter-Based Miniature Spectrometers: Spectrum Reconstruction Using Adaptive Regularization},
	volume = {11},
	issn = {1558-1748},
	url = {https://doi.org/10.1109/JSEN.2010.2103054},
	doi = {10.1109/JSEN.2010.2103054},
	shorttitle = {Filter-Based Miniature Spectrometers},
	pages = {1556--1563},
	number = {7},
	journal = {{IEEE} Sensors Journal},
	author = {Kurokawa, U. and Choi, B. I. and Chang, C.-C.},
	urldate = {2025-08-26},
	date = {2011-07},
    year = 2011
}

@article{yuan_wavelength-scale_2021,
	title = {A wavelength-scale black phosphorus spectrometer},
	volume = {15},
	issn = {1749-4885},
	doi = {10.1038/s41566-021-00787-x},
	pages = {601--607},
	journal = {Nature Photonics},
	author = {Yuan, S. and Naveh, D. and Watanabe, K. and Taniguchi, T. and Xia, F.},
	urldate = {2025-08-26},
	date = {2021-08-01},
	note = {{ADS} Bibcode: 2021NaPho..15..601Y},
        year = 2021
}

@article{zheng_deep_2021,
	title = {Deep plug-and-play priors for spectral snapshot compressive imaging},
	volume = {9},
	issn = {2327-9125},
	doi = {10.1364/PRJ.411745},
	pages = {B18},
	number = {2},
	journal = {Photonics Research},
	shortjournal = {Photon. Res.},
	author = {Zheng, S. and Liu, Y. and Meng, Z. and Qiao, M. and Tong, Z. and Yang, X. and Han, S. and Yuan, X.},
	urldate = {2025-08-26},
	date = {2021-02-01},
	langid = {english},
        year = 2021
}

@article{ma_robust_2024,
	title = {Robust spectral reconstruction algorithm enables quantum dot spectrometers with subnanometer spectral accuracy},
	volume = {3},
	issn = {2791-1519, 2791-1519},
	pages = {046009},
	number = {4},
	journal = {Advanced Photonics Nexus},
	shortjournal = {{APN}},
	author = {Ma, W. and Xue, Q. and Yang, Y. and Zhang, H. and Zhang, D. and Lan, X. and Gao, L. and Zhang, J. and Tang, J.},
	urldate = {2025-08-26},
	date = {2024-06},
	note = {Publisher: {SPIE}},
        year = 2024
	}

@inproceedings{song2023pseudoinverse,
  author    = {Song, J. and Vahdat, A. and Mardani, M. and Kautz, J.},
  title     = {Pseudoinverse-Guided Diffusion Models for Inverse Problems},
  booktitle = {Proc. International Conference on Learning Representations (ICLR)},
  year      = {2023}
}

@article{candes2006robust,
  title={Robust uncertainty principles: Exact signal reconstruction from highly incomplete frequency information},
  author={Cand{\`e}s, E. J. and Romberg, J. and Tao, T.},
  journal={IEEE Transactions on information theory},
  volume={52},
  number={2},
  pages={489--509},
  year={2006},
  publisher={IEEE}
}

@article{meerdink_ecostress_2019,
  title = {The {ECOSTRESS} spectral library version 1.0},
  volume = {230},
  issn = {0034-4257},
  doi = {https://doi.org/10.1016/j.rse.2019.05.015},
  author = {Meerdink, Susan K. and Hook, Simon J. and Roberts, Dar A. and Abbott, Elsa A.},
  date = {2019},
  journal = {Remote Sensing of Environment},
  year = {2019},
  month = sep,
  pages = {111196},
}

@article{zhang_snapshot_2024,
  author  = {Zhang, Haomin and Li, Quan and Zhao, Huijuan and Wang, Bowen and Gong, Jiaxing and Gao, Li},
  title   = {Snapshot computational spectroscopy enabled by deep learning},
  journal = {Nanophotonics},
  volume  = {13},
  number  = {22},
  pages   = {4159--4168},
  year    = {2024},
  doi     = {10.1515/nanoph-2024-0328},
}

@article{barata2012moore,
  title={The Moore--Penrose pseudoinverse: A tutorial review of the theory},
  author={Barata, J. C. A. and Hussein, M. S.},
  journal={Brazilian Journal of Physics},
  volume={42},
  number={1},
  pages={146--165},
  year={2012},
  publisher={Springer}
}
\vspace{12pt}
\end{document}